\documentclass[10pt,journal]{IEEEtran}
\usepackage{cite}
\usepackage{amsmath,amssymb,amsfonts}
\usepackage{graphicx}
\usepackage{xcolor}
\usepackage{booktabs}
\usepackage{array}
\usepackage{url}
\usepackage{hyperref}

\hypersetup{hidelinks}
\usepackage{amsmath}
\usepackage{bm} 
\usepackage{microtype}

\title{Image CL-SEC: Cross-Layer Semantic Error Correction for Image Communication}
\author{
Feifan~Zhang$^*$, 
Xuelin~Bai$^*$, 
Yirun~Wang$^*$, 
Yuyang~Du, 
Soung~Chang~Liew~\IEEEmembership{Life Fellow,~IEEE}

\thanks{F. Zhang, Y. Du, Y. Wang and S. C. Liew are with the Department of Information Engineering, The Chinese University of Hong Kong, HKSAR (e-mail: \{zf024, yrwang, yuydu, soung\}@ie.cuhk.edu.hk).}
\thanks{X. Bai is with the Department of Software, Chung-Ang University, Seoul, Republic of Korea. This work was conducted during X. Bai's internship at CUHK (email: baixuelin307@gmail.com)}
\thanks{$^*$F. Zhang, X. Bai, and Y. Wang contribute equally to this work.}
\vspace{-1.5em}
}

\begin{document}
\maketitle

\begin{abstract}
Semantic error correction (SEC) uses the semantic knowledge of pretrained models to complement conventional channel decoding. By combining physical-layer decoding information with application-layer semantic predictions, cross-layer SEC (CL-SEC) has demonstrated substantial performance gains in text transmission; however, its applicability to image transmission remains largely unexplored. This paper proposes Image CL-SEC, which, to the best of our knowledge, is the first extension of the CL-SEC framework to image transmission. Image CL-SEC uses channel-decoding information together with contextual predictions from a pretrained visual model to resolve residual uncertainty after forward error correction (FEC) decoding. In our implementation, images are represented as discrete visual tokens. For each token position selected for re-evaluation, the probability distribution produced by the channel decoder is combined with the distribution predicted by the visual model to identify the most likely transmitted token. The resulting decision therefore reflects both channel evidence and visual context. Experiments across multiple channel conditions show that, at low and moderate signal-to-noise ratios (SNRs), Image CL-SEC improves pixel-level fidelity, perceptual similarity, and token-recovery accuracy relative to either channel decoding or visual-model inference alone. Supplementary experiments using different pretrained visual models further demonstrate that these gains are not specific to a single model.
\end{abstract}
\section{Introduction}
\label{sec:introduction}

Advances in the contextual understanding and inference capabilities of language models (LMs) have generated growing interest in semantic error correction (SEC) as an emerging direction in communications systems. SEC leverages an LM's semantic understanding of transmitted content to assist error correction. Early studies used an LM at the application layer to correct residual text errors after conventional channel decoding \cite{SydneySEC1,SydneySEC2}. Subsequent work introduced cross-layer semantic error correction (CL-SEC) \cite{clsec} and demonstrated it on real-world hardware \cite{hardware}. Unlike earlier application-layer-only approaches, CL-SEC in \cite{clsec,hardware} jointly exploits (1) physical-layer information obtained from channel decoding and (2) application-layer semantic predictions generated by an LM, substantially improving error-correction performance. More recently, inspired by the success of LDPC and turbo codes, iterative CL-SEC \cite{ICL-SEC} was proposed as an iterative extension in which iterative message passing progressively refines decoding decisions. Iterative CL-SEC achieves state-of-the-art SEC performance, reducing the bit error rate (BER) by \textbf{more than four orders of magnitude} relative to a conventional LDPC baseline using belief-propagation (BP) decoding only.

Despite this rapid progress, SEC research to date \cite{SydneySEC1,SydneySEC2,clsec,ICL-SEC,SydneySEC3} has focused primarily on text transmission, leaving its application to non-text modalities largely unexplored. Meanwhile, emerging visual applications, such as augmented reality (AR) and virtual reality (VR), are placing increasingly stringent demands on the reliability of image transmission \cite{URLLC_overview,VrQos,ARsurvey}. Extending CL-SEC to the image domain is therefore an important step toward broadening its applicability beyond text and supporting reliability-critical visual applications.

Like their text-based counterparts, visual large models (VLMs) have demonstrated strong capabilities in image understanding and contextual inference \cite{VLM}, providing a foundation for extending SEC to the image domain. The central question addressed in this paper is \textit{how these capabilities can be effectively combined with physical-layer channel-decoding information to improve reliability.}

\begin{figure*}[!t]
\centering
\includegraphics[width=\textwidth]{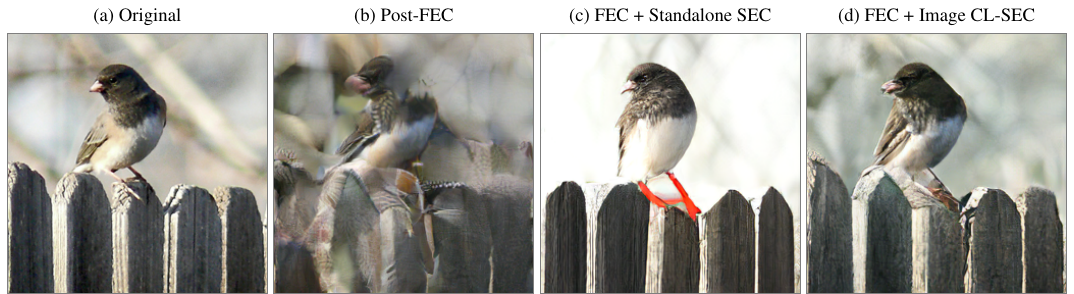}
\caption{Image recovery results over an AWGN channel at an SNR of 1 dB.
(a) Original image. (b) Reconstruction after FEC decoding. (c)
Reconstruction after FEC decoding and standalone SEC. (d) Reconstruction
after FEC decoding and Image CL-SEC.}
\label{fig:1}
\end{figure*}

To that end, we propose Image CL-SEC. Its central idea is to combine, within a Bayesian framework, probability estimates derived from physical-layer channel decoding with semantic likelihoods provided by an application-layer VLM. To see the benefits of this approach, Fig.~\ref{fig:1} presents an image recovery example from our experiments, with the original image shown in Fig.~\ref{fig:1}(a). After this image is transmitted over a noisy wireless channel and undergoes forward error correction (FEC) decoding at the receiver, residual errors cause noticeable local distortions in the reconstructed bird and background, as shown in Fig.~\ref{fig:1}(b). A standalone SEC method, following the application-layer approach as in an earlier text-based SEC study \cite{SydneySEC1}, uses a visual model to alleviate these distortions and produce a more coherent reconstruction. However, it incorrectly renders the bird's legs red, as shown in Fig.~\ref{fig:1}(c). This is because the retained image context may not provide enough information to determine the original color, leading the visual model to rely more heavily on visual patterns learned during pretraining. Its completion may therefore be semantically plausible (e.g., red bird legs) yet inconsistent with the original color. To determine which color is more likely to match the transmitted image, Image CL-SEC combines the visual model's contextual predictions at the affected locations with the corresponding soft information from the channel decoder that retains evidence about the transmitted content, thereby enabling faithful image recovery, as shown in Fig.~\ref{fig:1}(d).

The core contribution of this paper is the design and implementation of the Image CL-SEC framework, which, to the best of our knowledge, is the first realization of SEC for image transmission. Because representative visual models process images as visual tokens \cite{VQVAE,VLMtoken,MAGE,MaskGIT}, Image CL-SEC similarly represents each image as a sequence of discrete visual tokens. Fig.~\ref{fig:architecture} illustrates the overall framework. At the transmitter, Image CL-SEC encodes an image into discrete visual tokens representing local visual features and transmits their indices over the wireless channel. At the receiver, post-FEC tokens whose confidence falls below a predefined threshold are treated as unreliable and masked. Based on the surrounding context provided by the unmasked, higher-confidence tokens, a VLM estimates a probability distribution over the possible token values at each masked position. This distribution is then combined with the corresponding soft information from the channel decoder using Bayes' rule, and the token index with the highest posterior probability is selected as the final token estimate.

This paper not only establishes the feasibility of SEC for image transmission but also demonstrates its substantial performance gains through comprehensive experiments. Image CL-SEC consistently improves performance across different channel conditions, evaluation metrics, and pretrained VLMs, highlighting its generality as a framework. As a representative example, over an AWGN channel at an SNR of 0 dB with MAGE \cite{MAGE} as the backbone VLM, Image CL-SEC outperforms standalone SEC in both pixel-level fidelity and perceptual similarity, achieving an approximately 4.0 dB increase in peak signal-to-noise ratio (PSNR) and a 30.5\% reduction in learned perceptual image patch similarity (LPIPS) distance \cite{lpips}. Under the same setting, it reduces the token error rate (TER) from 89.6\% to 46.5\%, corresponding to a relative reduction of 48.1\%.

To facilitate broader participation and further research in this area, we also provide Image CL-SEC as a reproducible baseline. The implementation of the framework will be released as open-source software upon publication of this paper.

\section{System Architecture}
\label{sec:system-architecture}

\begin{figure*}[!t]
\centering
\includegraphics[width=\textwidth]{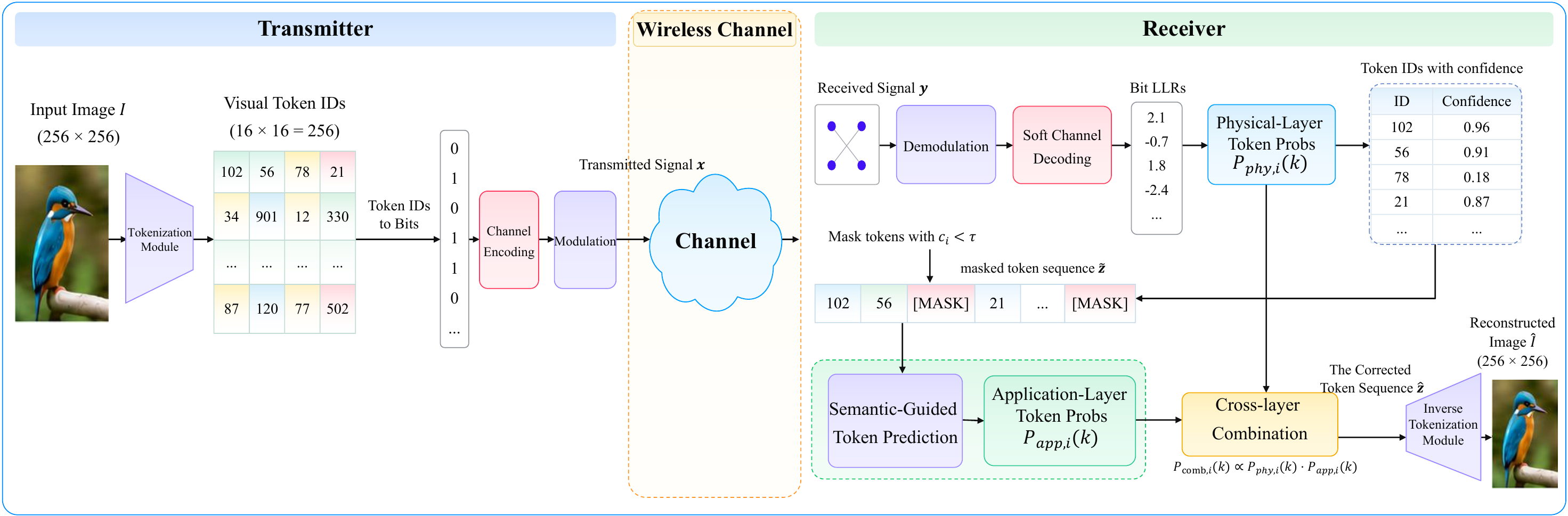}
\caption{Image CL-SEC architecture. The receiver masks initial
token estimates whose confidence falls below a predefined threshold and
re-estimates the tokens at the masked positions by combining probability
distributions from the physical-layer channel decoders and
application-layer VLM. The unmasked token estimates remain unchanged
during cross-layer processing.}
\label{fig:architecture}
\end{figure*}

Image CL-SEC is designed for image communication systems that represent images using discrete visual tokens, as illustrated in Fig.~\ref{fig:architecture}. At the transmitter, an image is converted into a sequence of token indices, each associated with an image patch. The token indices are then channel-encoded, modulated, and transmitted over the wireless channel. At the receiver, Image CL-SEC combines soft information from the channel decoder with contextual predictions from a VLM to estimate the transmitted tokens, from which the image is reconstructed using a visual decoder. This section provides an overview of the end-to-end process, while Sec.~\ref{sec:key-components} presents the implementation details of each module.

At the transmitter, the \emph{\textbf{Tokenization module}} converts the input image \(I \in \mathbb{R}^{H \times W \times 3}\) into a discrete visual tokens, where \(H\) and \(W\) denote the image height and width, respectively, and the number 3 denotes the RGB color channels. The token positions form a regular two-dimensional grid, with each position corresponding to an equal-sized, non-overlapping image region referred to as a patch. For example, the \(256 \times 256\) image in Fig.~\ref{fig:architecture} is represented by a \(16 \times 16\) token grid, so each token position corresponds to a \(16 \times 16\)-pixel patch. The visual encoder processes the entire image and produces a feature vector for each token position, which captures the information of the corresponding patch.

To convert these continuous feature vectors into discrete indices suitable for transmission, the transmitter applies codebook quantization. The codebook contains \(K\) candidate vectors (i.e., the tokens), and each vector extracted from the image is mapped to the index of its nearest codebook vector. The resulting visual token IDs form a two-dimensional grid. This grid is then flattened into a one-dimensional sequence in row-major order, i.e., each row is read from left to right, proceeding from the top row to the bottom row, yielding the visual token sequence

\begin{equation}
\mathbf{z}=E(I)=\left(z_1,z_2,\ldots,z_N\right),\quad z_i\in\mathcal{K},
\label{eq:token-sequence}
\end{equation}
where \(E\) denotes the visual encoding, quantization, and grid flattening, \(N\) is the number of tokens, and \(\mathcal{K} = \{0,\ldots,K-1\}\) is the visual codebook index set.

The transmitter represents each token ID as a \(B = \lceil \log_2 K \rceil\)-bit binary number and concatenates these representations to form a source bit sequence of length \(NB\), denoted by \(\mathbf{b}\). This source bit sequence is then channel-encoded and modulated to generate the symbols transmitted over the wireless channel. The \(j\)th received symbol is

\begin{equation}
y_j=h_jx_j+n_j.
\label{eq:channel-model}
\end{equation}
where \(x_j\) is the transmitted symbol, \(h_j\) is the channel coefficient, and \(n_j \sim \mathcal{CN}(0,\sigma_n^2)\) is additive complex Gaussian noise with power \(\sigma_n^2 = \mathbb{E}[|n_j|^2]\). For AWGN channels, we have \(h_j = 1\). For Rayleigh fading channels, the coefficients are independently drawn as \(h_j \sim \mathcal{CN}(0,1)\), giving unit average channel power, \(\mathbb{E}[|h_j|^2] = 1\). We evaluate Image CL-SEC under both channel models (see Sec.~\ref{subsec:experimental-setup} for the experimental settings).

At the receiver, the received symbols are first demodulated and channel-decoded to obtain bit estimates and associated soft information. Following the bit ordering used at the transmitter, the receiver estimates the most likely token ID for each group of \emph{B} bits, thereby forming an initial token sequence. The confidence of each token estimate is then compared with a predefined threshold. Token estimates whose confidence falls below the threshold are considered potentially unreliable and masked for subsequent reevaluation, whereas the remaining estimates are treated as reliable and provide visual context for the VLM to predict the tokens at the masked positions.

Assessing the confidence of each token estimate requires the decoder's soft information for its associated bits. The decoder expresses this information using log-likelihood ratios (LLR), where each LLR is the logarithm of the ratio between the probabilities that the corresponding source bit is 0 and 1 (see Eq.~\eqref{eq:bit-llr} in Sec.~\ref{subsec:physical-layer}). The \emph{\textbf{Physical-Layer Token Probs}} module in Fig.~\ref{fig:architecture} combines the bit-level LLRs associated with each token position to construct a probability distribution over all possible token IDs. Further details are provided in Sec.~\ref{subsec:physical-layer}. For the \(i\)th token position, \(P_{\mathrm{phy},i}(k)\) denotes the probability assigned to ID \(k \in \mathcal{K}\) based on the channel-decoding information. Let \(\hat z_i^{\mathrm{phy}}\) denote the ID of the most likely token (i.e., \(\hat z_i^{\mathrm{phy}} = \operatorname*{arg\,max}_{k} P_{\mathrm{phy},i}(k)\)). The confidence of the most likely token is

\begin{equation}
c_i=P_{\mathrm{phy},i}\!\left(\hat{z}_i^{\mathrm{phy}}\right).
\label{eq:token-confidence}
\end{equation}

Using these confidence scores, the receiver applies a pre-defined threshold \(\tau\) to determine which token estimates are retained as trusted context and which requires further processing. Specifically, if \(c_i \ge \tau\), the token estimate at that position is retained used as visual context. If \(c_i < \tau\), the estimate is regarded as unreliable, and the corresponding token is replaced with a special mask symbol that marks the position for subsequent re-evaluation. This process produces a partially masked token sequence in which the unmasked tokens provide trusted visual context, while the mask tokens indicate the positions for which the VLM generates contextual predictions.

The \emph{\textbf{Semantic-Guided Token Prediction}} module in Fig.~\ref{fig:architecture} uses a pretrained VLM to predict a token probability distribution for each masked position based on the context provided by the retained tokens. These distributions are defined over the same visual codebook and are represented by the \textbf{\emph{Application-Layer Token Probs}} module in Fig.~\ref{fig:architecture}. For each masked position \(i\), let \(P_{\mathrm{app},i}(k)\) denote the predicted probability of token ID \(k\) , referred to as the application-layer token probability. Its computation is detailed in \eqref{eq:application-probability} in Sec.~\ref{subsec:application-layer}.

For each masked position, the receiver then combines the application-layer predictions with the physical-layer probabilities computed before masking. As illustrated in Fig.~\ref{fig:architecture}, the physical-layer probabilities are passed directly from the \emph{\textbf{Physical-Layer Token Probs}} module to the \emph{\textbf{Cross-Layer Combination}} module. For each masked position, this module approximates Bayesian fusion by multiplying the corresponding candidate-token probabilities from the two layers elementwise and normalizing the resulting products to obtain an approximate posterior distribution \(P_{\mathrm{comb},i}(k)\) (see Sec.~\ref{subsec:cross-layer} for details). The token ID with the highest posterior probability is selected as the final token estimate result. The token IDs at unmasked positions remain unchanged.

Finally, the corrected token sequence \(\mathbf{\hat z}\), as shown in Fig.~\ref{fig:architecture}, is passed to the \emph{\textbf{Inverse Tokenization}} module. Following the ordering rule used at the transmitter, this module reshapes the one-dimensional sequence into a two-dimensional token grid, retrieves the corresponding feature vectors through codebook lookup, and applies the visual decoder to reconstruct the image:

\begin{equation}
\hat{I}=D\!\left(\mathbf{\hat{z}}\right),
\label{eq:image-reconstruction}
\end{equation}
where \(D\) denotes sequence reshaping, codebook lookup, and visual decoding, and \(\hat I\) is the image reconstructed at the receiver.

\section{Key Components of Image CL-SEC}
\label{sec:key-components}

This section details the construction of physical-layer and application-layer token probability distributions and explains how they are combined to perform cross-layer token estimation.

\subsection{Physical-Layer Token Probability Distribution}
\label{subsec:physical-layer}

The physical layer constructs a probability distribution over the possible token IDs at each position to assess the reliability of the initial token estimate and facilitate subsequent cross-layer probability combination. Since each token ID corresponds to a unique bit pattern, the receiver can estimate the probability of each candidate ID using the soft information provided by the channel decoder for the associated bits. The computation consists of two steps: (1) converting bit LLRs into bit-value probabilities and (2) combining these probabilities into a token-level probability distribution.

In step (1), the receiver converts the LLR of each bit into the probabilities that the bit takes the values 0 and 1. For the \emph{i}-th token, let \(b_{i,m} \in \{0,1\}\) denote the \emph{m}-th source bit, where \(m = 1,\ldots,B\). The LLR output by the channel decoder for this bit, denoted by \(L_{i,m}\) , is

\begin{equation}
L_{i,m}=\ln\frac{\Pr\!\left(b_{i,m}=0\mid y\right)}{\Pr\!\left(b_{i,m}=1\mid y\right)},
\label{eq:bit-llr}
\end{equation}
where \(y\) is the received symbol sequence. Since the probabilities of ``bit 0'' and ``bit 1'' sum to one, solving \eqref{eq:bit-llr} yields

\begin{equation}
\begin{aligned}
p_{i,m}(a)
&=\Pr\!\left(b_{i,m}=a\mid y\right) \\
&=\frac{1}{1+\exp\!\left((2a-1)L_{i,m}\right)},
\quad a\in\{0,1\}.
\end{aligned}
\label{eq:bit-probability}
\end{equation}

In step (2), the receiver combines these bit probabilities according to the binary representation of each candidate token ID. For each candidate ID \(k \in \mathcal{K}\), let \(\beta_m(k) \in \{0,1\}\) denote the \emph{m}-th bit in its binary representation. If ID \(k\) was transmitted at position \(i\), its associated bits would be \(\beta_1(k),\ldots,\beta_B(k)\), respectively. To estimate the probability of each candidate ID, we draw on the combination rule in belief propagation (BP), which multiplies a local prior by incoming messages. For candidate token ID \(k\), \(Q(k)\) denotes its prior probability, while \(p_{i,m}(\beta_m(k))\) represents the support provided by channel decoding for its \(m\)-th bit value. Using these bit probabilities as soft information and combining them with the token prior, we obtain

\begin{equation}
P_{\mathrm{phy},i}(k)\propto Q(k)\prod_{m=1}^{B}p_{i,m}\!\left(\beta_m(k)\right),\quad k\in\mathcal{K}.
\label{eq:token-probability-prior}
\end{equation}

We do not exploit prior token statistics and therefore set \(Q(k) = 1/K\), treating all candidate IDs as equally likely a priori. The codebooks used in this work satisfy \(K = 2^B\), so every \(B\)-bit pattern corresponds to a valid ID. With \(Q(k) = 1/K\), the unnormalized expression in Eq.~\eqref{eq:token-probability-prior} therefore sums to \(1/K\). Normalization cancels the uniform prior factor, yielding the physical-layer token probability distribution\footnote{This product form does not explicitly account for possible dependencies among bits conditioned on the received information. When such dependencies exist, the resulting token probabilities are approximate estimates.}

\begin{equation}
P_{\mathrm{phy},i}(k)=\prod_{m=1}^{B}p_{i,m}\!\left(\beta_m(k)\right),\quad k\in\mathcal{K}.
\label{eq:token-probability}
\end{equation}

The receiver uses this distribution to compute the confidence of the initial token estimate at each position, as defined in \eqref{eq:token-confidence}. For positions identified as unreliable and subsequently masked, the complete physical-layer probability distributions are retained for cross-layer combination.

\subsection{Application-Layer Token Probability Distribution}
\label{subsec:application-layer}

The application layer uses a pretrained VLM to estimate a token probability distribution for each masked position selected for re-evaluation, conditioned on the visual context provided by the retained tokens. During pretraining, a subset of image tokens is masked, and the model learns to predict their original IDs from the remaining, unmasked tokens \cite{MAGE,MaskGIT}. Through this masked-token prediction objective, the VLM learns contextual relationships among image regions and can therefore infer missing visual content from a partially masked token sequence.

As described in Section~\ref{sec:system-architecture}, for the input to the VLM, the receiver replaces token estimates whose confidence is below the predefined threshold \(\tau\) with masks and retains the remaining estimates as context. Let \(\mathcal{M}\) denote the set of masked positions and \(z\) the masked token sequence. They are defined as

{\small
\begin{equation}
\mathcal{M}=\left\{i\in\{1,\ldots,N\}:c_i<\tau\right\}, z_i=\begin{cases}[\mathrm{MASK}],&i\in\mathcal{M},\\\hat{z}_i^{\mathrm{phy}},&i\notin\mathcal{M},\end{cases}
\label{eq:masked-sequence}
\end{equation}
}
where \([\mathrm{MASK}]\) is a special symbol marking a position to be predicted. The masked sequence preserves the original ordering, so each sequence position still corresponds to its original image region.

Denote the pretrained visual model by \(G_\theta\), where \(\theta\) denotes its pretrained parameters. Given the masked sequence \(z\), the model outputs \(K\) candidate scores for each masked position based on the context provided by the retained tokens. At position \(i\), the score for candidate \(k\) is denoted by \(s_i(k)\), reflecting the model's relative preference for that ID under the given context. The model then applies SoftMax to these scores to obtain predicted probabilities, which we use as the application-layer token probability distribution:

\begin{equation}
P_{\mathrm{app},i}(k)=\frac{\exp\!\left(s_i(k)\right)}{\sum_{u\in\mathcal{K}}\exp\!\left(s_i(u)\right)},\quad i\in\mathcal{M}.
\label{eq:application-probability}
\end{equation}
where \(P_{\mathrm{app},i}(k)\) is the probability predicted by the model that the token ID at position \(i\) is \(k\), given the current image context.

For subsequent product combination, the physical-layer and application-layer probability distributions must be defined over the same codebook index set \(\mathcal{K}\), with each token ID referring to the same codebook vector in both layers. Therefore, the pretrained visual model must use a token-ID mapping consistent with the tokenizer and codebook employed at the transmitter. Sec.~\ref{subsec:experimental-setup} describes the prediction models and their configurations.

\subsection{Cross-Layer Token Evaluation}
\label{subsec:cross-layer}

Sections~\ref{subsec:physical-layer} and~\ref{subsec:application-layer} construct the physical-layer and application-layer token probability distributions, respectively. This subsection describes how these distributions are combined within a Bayesian framework to re-estimate the tokens at masked positions.

Text-based CL-SEC \cite{clsec} jointly uses channel-decoding information and a language model's contextual predictions for word correction. The former provides signal-based support for candidate words, while the latter provides their predicted probabilities given the current context. The method performs Bayesian combination in product form by multiplying and normalizing the two layers' probabilities for each candidate word to make correction decisions. Text-based CL-SEC \cite{clsec} explains the rationale for this product form from the perspective of message combination in belief propagation (BP). We refer readers to Sec. 3.6 and Appendix A of \cite{clsec} for details of the corresponding analysis and probability derivations. We follow this approach for visual token estimation by combining, in product form, the token probabilities derived from physical-layer channel decoding with those predicted by the visual model from the image context.

Specifically, for each masked position \(i \in \mathcal{M}\), the receiver multiplies the physical-layer probability \(P_{\mathrm{phy},i}(k)\) and the application-layer probability \(P_{\mathrm{app},i}(k)\) for each candidate ID \(k\). The products are then normalized by their sum over all candidate IDs to obtain the approximate posterior distribution\footnote{Since the two sources of information may still be statistically dependent, conditional independence is treated here as an approximation.}

\begin{equation}
P_{\mathrm{comb},i}(k)=\frac{P_{\mathrm{phy},i}(k)P_{\mathrm{app},i}(k)}{\sum_{u\in\mathcal{K}}P_{\mathrm{phy},i}(u)P_{\mathrm{app},i}(u)},\quad i\in\mathcal{M}.
\label{eq:combined-probability}
\end{equation}

Here, \(P_{\mathrm{comb},i}(k)\) denotes the approximate posterior probability of candidate ID \(k\) at position \(i\). It reflects both the support from transmission information and the candidate token's compatibility with the image context, providing a basis for token estimation.

Using the approximate posterior distribution in \eqref{eq:combined-probability}, the receiver selects the highest-probability ID at each masked position while retaining the initial physical-layer estimates at unmasked positions. The estimated token sequence is therefore given by

\begin{equation}
\hat{z}_i=\begin{cases}\displaystyle\arg\max_{k\in\mathcal{K}}P_{\mathrm{comb},i}(k),&i\in\mathcal{M},\\\hat{z}_i^{\mathrm{phy}},&i\notin\mathcal{M}.\end{cases}
\label{eq:token-estimate}
\end{equation}

The resulting token sequence \(\mathbf{\hat z}\) is then passed to the Inverse Tokenization module to reconstruct the image (see Eq.~\eqref{eq:image-reconstruction} for the reconstruction mapping).

\section{Experimental Results}
\label{sec:experimental-results}

This section evaluates Image CL-SEC under AWGN and Rayleigh fading channels. We will examine the system's token recovery performance and the image reconstruction quality. And this followed by a visual case study for illustration. Supplementary experiments with another pretrained visual model are provided to demonstrate the universality of the proposed framework (see Appendix~\ref{sec:supplementary}).

\subsection{Experimental Setup}
\label{subsec:experimental-setup}

\textbf{Visual models.} We use MAGE \cite{MAGE} as the pretrained visual model at the application layer, together with its corresponding pretrained VQGAN \cite{VQGAN} for image tokenization and decoding. Through masked image modeling, MAGE learns contextual relationships within images and predicts missing tokens from the retained content. Each \(256 \times 256\) image is represented by a \(16 \times 16\) grid of \(N = 256\) tokens. The codebook contains \(K = 1024\) entries, and each token ID is represented by \(B = 10\) bits, yielding 2560 source bits per image before channel coding. To examine whether the benefits of image CL-SEC extend to other pretrained visual models, we also conduct supplementary experiments using MaskGIT \cite{MaskGIT} (Appendix~\ref{sec:supplementary}).

\textbf{Dataset.} We randomly select 5000 images from ImageNet-1K \cite{imagenet} to form the test dataset and resize them to \(256 \times 256\). All experiments use the same test dataset.

\textbf{Communication and masking settings.} We use a rate-1/2 convolutional code with constraint length 3 and generator polynomials \((7,5)_8\). The encoder starts in the zero state, and two zero tail bits are appended to each input sequence to terminate the trellis in the zero state. The encoded bit sequence is randomly interleaved before QPSK modulation. At the receiver, the coded-bit LLRs obtained from demodulation are deinterleaved and then passed to a log-domain BCJR (log-MAP) decoder. The resulting a posteriori LLRs for the source bits are used to construct the physical-layer token probability distributions.

The experiments use the AWGN and Rayleigh fading models defined in Section~\ref{sec:system-architecture}. The Rayleigh fading experiments adopt a fast-fading setting in which the channel coefficients are independent across transmitted symbols. For each SNR setting, the additive noise is independent of the channel coefficients and has a fixed variance \(N_0 = \mathbb{E}[|n_j|^2]\). Under the channel model in Eq.~\eqref{eq:channel-model}, with the average symbol energy normalized to unity, the average SNR in dB is \(10\log_{10}(1/N_0)\), while the instantaneous received SNR in linear units is \(\frac{|h_j|^2}{N_0}\). For the Rayleigh fading experiments, the receiver assumes perfect channel state information (CSI) and performs zero-forcing equalization before demodulation, implemented as \(\tilde{x}_j = \frac{y_jh_j^*}{|h_j|^2+\epsilon}\), where \(\epsilon =  10^{-8}\) ensures numerical stability. Both channel experiments use a fixed confidence threshold of \(\tau = 0.7\).

\textbf{Evaluation metrics.} We use token error rate (TER), peak signal-to-noise ratio (PSNR), and learned perceptual image patch similarity (LPIPS) \cite{lpips} to evaluate token recovery, pixel-level fidelity, and perceptual similarity, respectively.

TER is evaluated over all tokens in all images in the test dataset and measures the fraction of final token IDs that differ from the original transmitted IDs. For a test dataset containing \(N_{\mathrm{img}}\) images with \(N\) tokens each, TER is defined as

\begin{equation}
\mathrm{TER}=\frac{1}{N_{\mathrm{img}}N}\sum_{d=1}^{N_{\mathrm{img}}}\sum_{i=1}^{N}\mathbf{1}\!\left[\hat{z}_i^{(d)}\ne z_i^{(d)}\right].
\label{eq:ter}
\end{equation}
Here, \(z_i^{(d)}\) and \(\hat z_i^{(d)}\) denote the transmitted and final estimated IDs at position \(i\) in image \(d\), respectively. The indicator \(\mathbf{1}[\cdot]\) equals 1 when the IDs differ and 0 otherwise.

For an original image \(I\) and its reconstruction \(\hat I\), PSNR is calculated as

\begin{equation}
\mathrm{PSNR}(I,\hat{I})=10\log_{10}\!\left(\frac{255^2\cdot3HW}{\left\|I-\hat{I}\right\|_F^2}\right)\,\mathrm{dB}.
\label{eq:psnr}
\end{equation}
Here, each image contains \(H \times W\) pixels with three RGB channels and pixel values in \([0,255]\). The term \(\|I-\hat I\|_F^2\) sums the squared errors over all pixels and color channels. Higher PSNR indicates lower pixel-level reconstruction error.

LPIPS measures perceptual distance through differences in visual features related to image structure and texture. Using features extracted by a pretrained perceptual network, it is calculated as

\begin{equation}
\mathrm{LPIPS}(I,\hat{I})=\sum_{\ell}\frac{1}{H_\ell W_\ell}\left\|w_\ell\odot\left[\bar{F}_\ell(I)-\bar{F}_\ell(\hat{I})\right]\right\|_F^2.
\label{eq:lpips}
\end{equation}
Here, \(\bar F_\ell(I)\) denotes the features at layer \(\ell\), normalized to unit \(L_2\) norm across channels at each spatial position. \(H_\ell\) and \(W_\ell\) are the spatial dimensions of the feature map, and \(w_\ell\) contains the learned channel weights applied at every spatial position. The symbol \(\odot\) denotes elementwise multiplication, and the squared norm sums over all spatial positions and channels. Lower LPIPS indicates greater perceptual similarity. PSNR and LPIPS are calculated for each image and then averaged over the test dataset.

\textbf{Compared methods.} The main experiments compare the following three methods. Both Standalone SEC and Image CL-SEC use MAGE as the application-layer visual model.

\begin{quote}
\textbf{(1) PHY.} Images are reconstructed directly from the
channel-decoded token sequence without application-layer token
reevaluation.

\textbf{(2) Standalone SEC.} The visual model predicts masked tokens from the retained context, and each masked position is replaced with the ID having the highest application-layer probability.

\textbf{(3) Image CL-SEC.} Physical-layer and application-layer token
probabilities are combined in product form, and the ID with the highest
posterior probability is selected for each masked position.
\end{quote}

Standalone SEC and Image CL-SEC use the same initial token estimates and mask positions, with unmasked positions remaining unchanged. The two methods differ in whether physical-layer probabilities are used to select the final estimated token IDs, allowing the comparison to assess the contribution of cross-layer probability combination.

\subsection{Token Recovery Performance}
\label{subsec:token-recovery}

\begin{figure}[!htbp]
\centering
\includegraphics[width=0.49\textwidth]{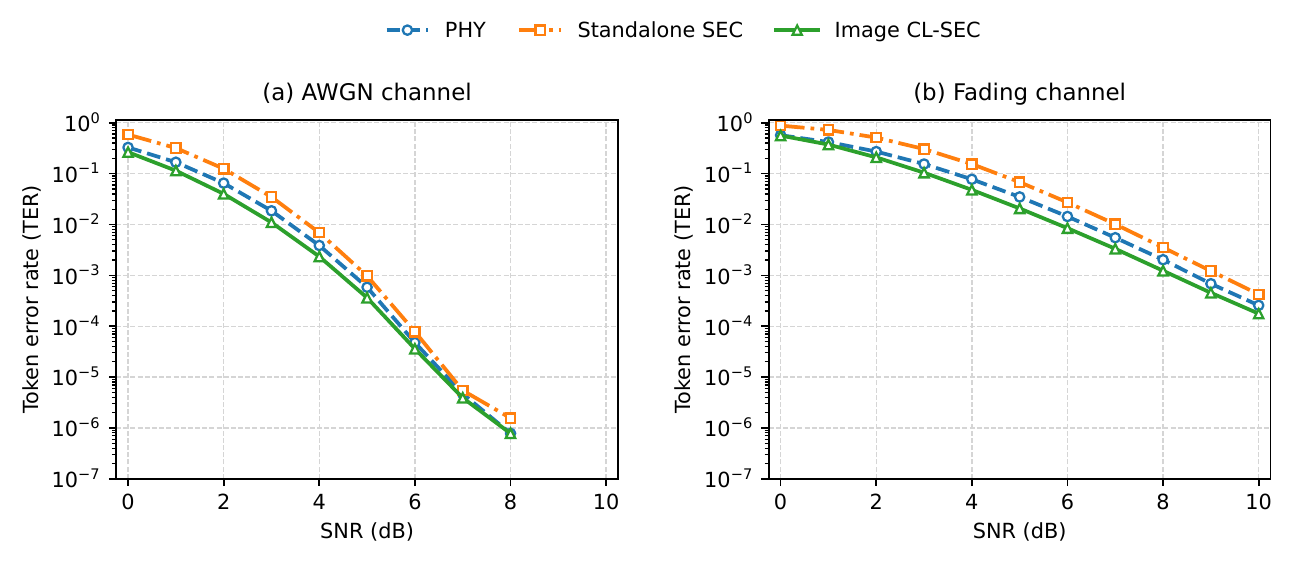}
\caption{Token error rates of the three methods under AWGN and Rayleigh
fading channels. Standalone SEC and Image CL-SEC use MAGE as the
application-layer visual model. TER is evaluated over all tokens in the
test dataset and plotted on a logarithmic scale. The confidence
threshold is
\(\tau = 0.7\).
Under the AWGN channel, all three methods yield zero observed token
errors at 9 and 10 dB. These points are omitted from the logarithmic
plot.}
\label{fig:3}
\end{figure}

Fig.~\ref{fig:3} compares TER under the two channel models. At each SNR, the evaluation includes all \(5000 \times 256 = 1{,}280{,}000\) tokens in the test dataset, rather than only masked positions. Lower TER indicates more accurate recovery of the original tokens.

TER decreases with increasing SNR for all three methods. Under both channel models, Image CL-SEC generally achieves lower TER than PHY and Standalone SEC, with particularly clear gains at low and moderate SNRs. In contrast, Standalone SEC has higher TER than PHY at some SNRs. This indicates that relying solely on the visual model's contextual semantic predictions does not necessarily recover the original tokens more accurately, whereas incorporating channel-decoding information helps improve token-recovery accuracy. At high SNRs, all methods have low TER, but improvements from cross-layer combination remain observable under the fading channel.

We note that changes in TER do not fully reflect image reconstruction quality, so the practical benefits of Image CL-SEC also need to be evaluated through end-to-end reconstruction results -- and this presented in Subsection~\ref{subsec:case-study} below.

\subsection{Image Reconstruction Quality}
\label{subsec:image-quality}

\begin{figure}[!htbp]
\centering
\includegraphics[width=0.49\textwidth]{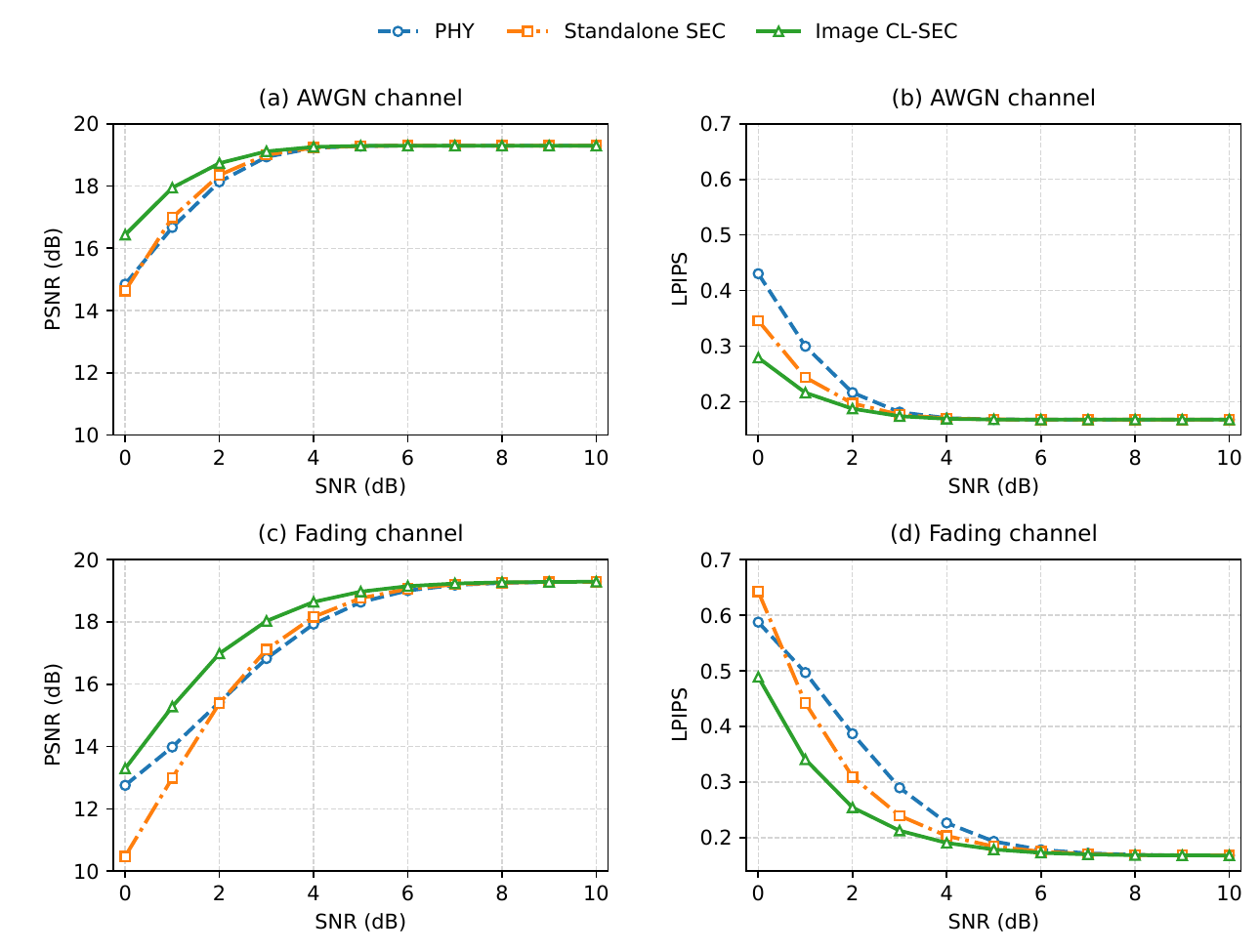}
\caption{Image reconstruction quality of the three methods under AWGN and Rayleigh fading channels. Standalone SEC and Image CL-SEC use MAGE as the application-layer visual model. The top row shows AWGN results over 0--10 dB, and the bottom row shows Rayleigh fading results over 0--20 dB. The left and right columns report PSNR and LPIPS, respectively. The confidence threshold is
\(\tau = 0.7\).}
\label{fig:4}
\end{figure}

\begin{figure*}[!t]
\centering
\includegraphics[width=0.6\textwidth]{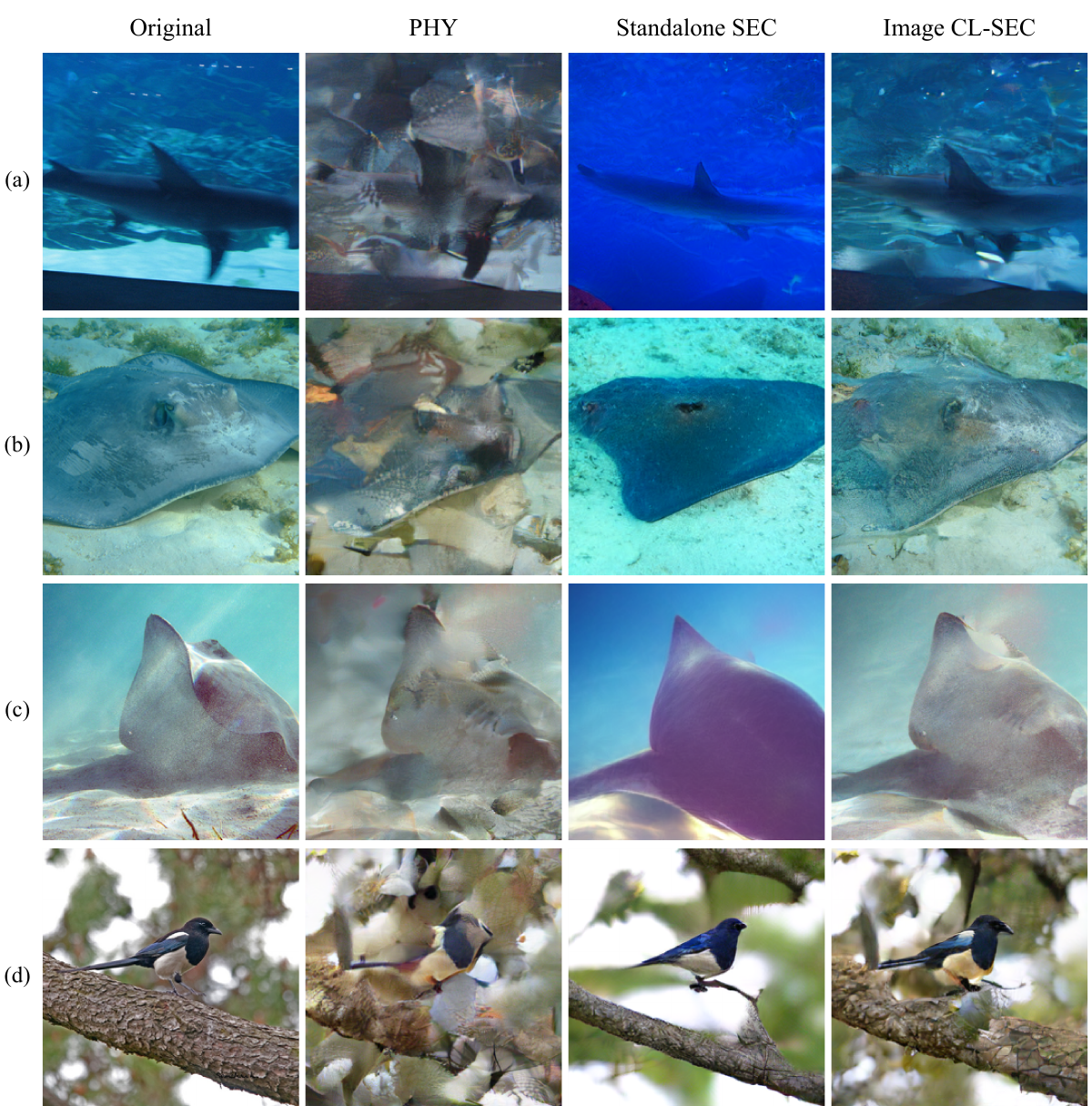}
\caption{Comparison of reconstructed images over an AWGN channel at an
SNR of 1 dB. Rows (a)--(d) show four cases. From left to right, the
columns show the original images and the images reconstructed using PHY,
Standalone SEC, and Image CL-SEC, respectively. Standalone SEC and Image
CL-SEC both use MAGE as the application-layer visual model}
\label{fig:case}
\end{figure*} 

Having evaluated token recovery performance, we next examine the quality of images reconstructed from the estimated token sequences. Fig.~\ref{fig:4} presents the PSNR and LPIPS results under both channel models. The results show that Image CL-SEC achieves the highest PSNR and lowest LPIPS among the three compared methods under AWGN and Rayleigh fading channels.

Together with the token recovery results in Fig.~\ref{fig:3}, these findings show that incorporating physical-layer channel-decoding information into application-layer token estimation does help recover the original tokens more accurately and improves end-to-end image reconstruction quality.

\subsection{Case Study}
\label{subsec:case-study}

To visually illustrate the image recovery performance, Fig.~\ref{fig:case} compares image reconstruction results for four cases over an AWGN channel at an SNR of 1 dB.

In Fig.~\ref{fig:case}, the PHY reconstructions exhibit noticeable distortions in local textures and object shapes. Standalone SEC alleviates these distortions and produces more coherent object appearances, but its completed content may still deviate from the original images. Figure~\ref{fig:case}(b) is a typical example of such deviations. Although Standalone SEC produces a more coherent appearance of the fish, its outline differs from that in the original image. In Figs.~\ref{fig:case}(a) and~\ref{fig:case}(c), Standalone SEC makes the underwater scene appear overly blue and changes the fish's color from gray to purple, respectively. It also alters the bird's color and background appearance in Fig.~\ref{fig:case}(d). In contrast, Image CL-SEC produces object appearances and backgrounds that more closely resemble those in the original images.

These cases show that Image CL-SEC helps reduce visually plausible semantic completions that deviate from the original content, making the reconstructed images more faithful to the originals.

\section{Conclusion}
\label{sec:conclusion}

This paper proposed Image CL-SEC, a cross-layer framework that helps resolve residual uncertainty after channel decoding to improve image recovery. Image CL-SEC combines token probability distributions derived from a physical-layer channel decoder and the application-layer predictions from a pretrained visual model, thereby jointly exploiting transmission evidence and visual context for token-level error correction. Experiments over AWGN and Rayleigh fading channels showed that, compared with channel decoding alone and standalone SEC, Image CL-SEC improved the pixel-level fidelity and perceptual similarity of the reconstructed images, as indicated by higher PSNR and lower LPIPS, while also increasing the token recovery accuracy. These results demonstrate the effectiveness of cross-layer probability combination in reducing image distortion caused by transmission errors. Additional experiments with different pretrained visual models further showed that the performance gains are not specific to a single model. Future work will investigate feedback from application-layer semantic prediction to physical-layer channel decoding, including the use of semantic predictions to assist subsequent decoding iterations and the potential of iterative cross-layer correction to further improve token recovery accuracy and image reconstruction quality.

\appendices

\setcounter{figure}{0}
\renewcommand{\thefigure}{A\arabic{figure}}
\renewcommand{\theHfigure}{appendix.A.\arabic{figure}}

\section{Supplementary Results with Different Visual Models}
\label{sec:supplementary}

To verify that the advantages of Image CL-SEC are not unique to MAGE,
this appendix provides supplementary experiments using MaskGIT as an
alternative pretrained visual model. The resulting data indicate that
the benefits indeed generalize to other frameworks.

\subsection{Supplementary Experimental Setup}
\label{subsec:supplementary-setup}

We compare three methods: PHY, Standalone SEC with MaskGIT, and Image CL-SEC with MaskGIT. The test dataset, settings of the wireless link, and settings of the masking policy are the same as in the main experiments, including a rate-\(1/2\) convolutional code, QPSK modulation, and a confidence threshold of \(\tau = 0.7\).

For each visual model, Standalone SEC and Image CL-SEC use identical initial token estimates and mask positions, differing only in whether physical-layer probabilities are used for token estimation. We use TER, PSNR, and LPIPS as evaluation metrics (see Sec.~\ref{subsec:experimental-setup} for their definitions).

\subsection{Token Recovery Performance}
\label{subsec:supplementary-token-recovery}

\begin{figure}[!htbp]
\centering
\includegraphics[width=\columnwidth]
{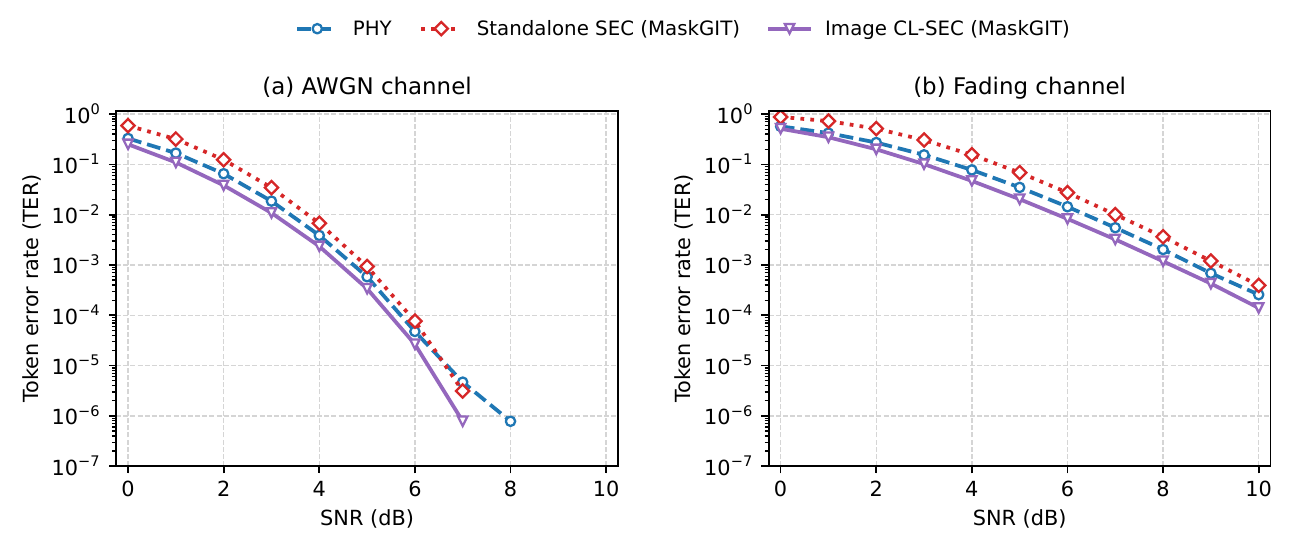}
\caption{Token error rates of the three methods under (a) AWGN and (b)
Rayleigh fading channels. TER is evaluated over all tokens in the test
dataset and plotted on a logarithmic scale. Points with zero observed
token errors are omitted. The confidence threshold is
\(\tau = 0.7\).}
\label{fig:A1}
\end{figure}

Fig.~\ref{fig:A1} compares the TER of the three methods under both channel models. With MaskGIT, Image CL-SEC improves token recovery performance over PHY and the corresponding Standalone SEC baseline. Together with the MAGE results in Sec.~\ref{subsec:token-recovery}, these results show that the benefits of incorporating channel-decoding information into application-layer token estimation are not limited to a single visual model.

\subsection{Image Reconstruction Quality}
\label{subsec:supplementary-image-quality}

\begin{figure}[!htbp]
\centering
\includegraphics[width=\columnwidth]
{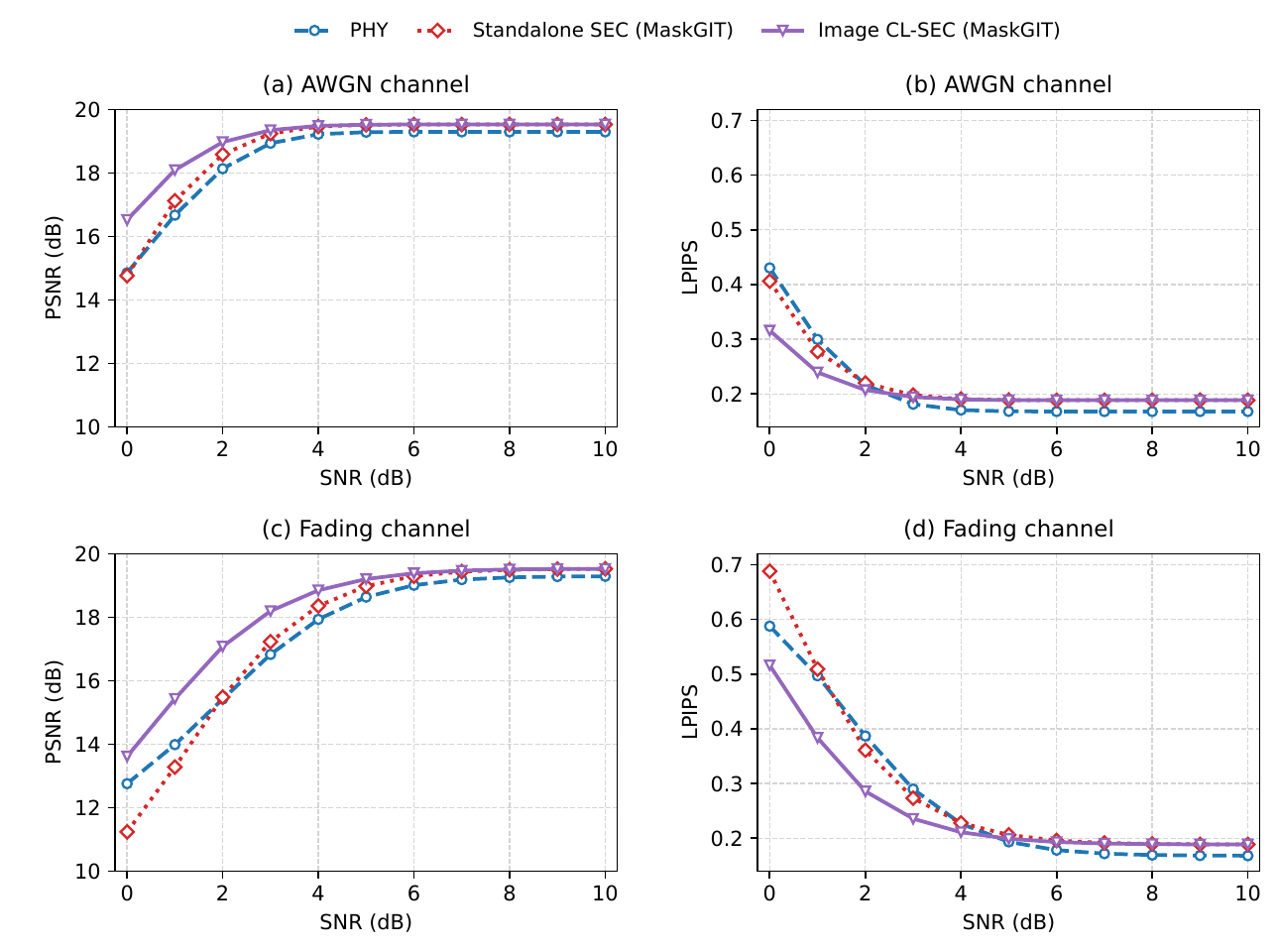}
\caption{Image reconstruction quality of the three methods under AWGN
and Rayleigh fading channels. The top and bottom rows show AWGN and
Rayleigh fading results, respectively. The left and right columns report
PSNR and LPIPS, respectively. The confidence threshold is
\(\tau = 0.7\).}
\label{fig:A2}
\end{figure}

Fig.~\ref{fig:A2} further compares PSNR and LPIPS for the three methods. With MaskGIT, Image CL-SEC also improves PSNR and reduces LPIPS relative to PHY and the corresponding Standalone SEC baseline, consistent with the gains observed using MAGE in Fig.~\ref{fig:4}. Together with Fig.~\ref{fig:A1}, these results show that cross-layer combination improves both original-token recovery and end-to-end image reconstruction with different visual models.

\bibliographystyle{IEEEtran}
\bibliography{references}

\end{document}